\documentclass[aps,prd,preprint,eqsecnum,tightenlines,floats,
groupedaddress,showpacs,superscriptaddress]{revtex4-1}
\usepackage{amssymb} 
\usepackage{hyperref} 
\usepackage{graphicx}
\usepackage{subfigure}
\usepackage{amsmath}
\usepackage{xcolor}
\usepackage{subfigure}

\def\beq{\begin{equation}}
\def\eeq{\end{equation}}

\newcommand{\p}{\vec{p}}

\newcommand{\N}{\mathcal{N}}
\renewcommand{\phi}{\varphi}
\renewcommand{\t}{t_{\rm cl}}

\begin{document}


\date{\today}

\title{Duration of Classicality of an inhomogeneous quantum field \\ with repulsive contact self-interactions}

\author{Ariel Arza}
\affiliation{Department of Physics, University of Florida, Gainesville, FL 32611, USA}
\author{Sankha S. Chakrabarty}
\affiliation{Department of Physics, University of Florida, Gainesville, FL 32611, USA}
\author{Seishi Enomoto}
\affiliation{School of Physics, Sun Yat-Sen University, Guangzhou 510275, China}
\affiliation{Department of Physics, University of Florida, Gainesville, FL 32611, USA}
\affiliation{Theory Center, High Energy Accelerator Research Organization (KEK),Tsukuba, Ibaraki 305-0801, Japan}
\author{Yaqi Han}
\affiliation{Department of Physics, University of Florida, Gainesville, FL 32611, USA}
\author{Elisa Todarello}
\affiliation{Department of Physics, University of Florida, Gainesville, FL 32611, USA}
\affiliation{Institut f\"ur Kernphysik, Karlsruhe Institute of Technology (KIT),\\ 76021 Karlsruhe, Germany}

\begin{abstract}

Quantum fields with large degeneracy are often approximated as classical fields. Here, we show how quantum and classical evolution of a highly degenerate quantum field with repulsive contact self-interactions differ from each other. Initially, the field is taken to be homogeneous except for a small plane wave perturbations in only one mode. In quantum field theory, modes satisfying both momentum and energy conservation of the quasi-particles, grow exponentially with time. However, in the classical field approximation, the system is stable. We calculate the time scale after which the classical field description becomes invalid.


\end{abstract}

\maketitle

\section{Introduction}
The duration of classicality $\t$ is the longest time for which the dynamics of a quantum system can be accurately approximated by the dynamics of an analogous classical system. 
Consider the following generic Hamiltonian
\beq
H = \sum_j \omega_j a_j^\dagger a_j + \sum_{jklm} \frac{1}{4}\Lambda_{jk}^{lm}\ a_j^\dagger a_k^\dagger a_l a_m  \enspace,
\label{1:hamiltonian_generic}
\eeq
where $[a_i, a_j^\dagger] = \delta_{ij}$. For example, we can think of the indices $j, k, l, m$  as labels for the  modes of a self-interacting Bosonic quantum field. The time evolution of this system is determined by the Heisenberg equations of motion
\beq
i\dot{a}_j = \omega_j a_j + \sum_{klm} \frac{1}{2} \Lambda_{jk}^{lm}\ a_k^\dagger a_l a_m  \enspace.\label{1:heisenberg}
\eeq
The analogous classical system is defined as the one that obeys the equations
\beq
i\dot{A}_j = \omega_j A_j + \sum_{klm} \frac{1}{2} \Lambda_{jk}^{lm}\ A_k^* A_l A_m  \enspace. \label{1:eom_class}
\eeq
Equations~\eqref{1:heisenberg} and~\eqref{1:eom_class} have the same structure and the same coefficients, the only difference being that each $A_j$ is a complex function, rather than an operator. 

Under what circumstances, and for how long, will Equations~\eqref{1:heisenberg} and~\eqref{1:eom_class} make similar predictions? Let us focus on the expectation value of the occupation number operators $\langle\N_j(t)\rangle =\langle\Psi\rvert a_j^\dagger(t) a_j(t)\lvert \Psi\rangle$ and compare them to their classical counterparts $N_j(t) = A_j^*(t)A_j(t)$.
The answer to the question above of course depends on the state of the quantum system $\lvert \Psi \rangle$. One might think that
a sufficient requirement for the classical and quantum evolutions to resemble each other is the occupation of the quantum oscillators being high, i.e. $\langle\N_j(t)\rangle \gg 1$. If that were true,  one could choose initial conditions such that $\langle\N_j(0)\rangle \approx N_j(0)$, and trust that the quantum and classical equations will make similar predictions for an indefinitely long time, up to corrections of order  $1/\N$, $\N$ being a typical value of $\langle\N_j(t)\rangle$. Perhaps, this assumption stems from the notion that a highly occupied quantum harmonic oscillator behaves as a classical harmonic oscillator, up to small corrections. 
However,  the presence of interactions plays a crucial role.

As was shown in Ref.~\cite{Sikivie:2016enz}, even in the high occupancy regime,  the classical approximation for a system of interacting quantum harmonic oscillators in general becomes invalid after a time $\t$, which is at most equal to the thermalization time scale of the quantum system $\tau$ multiplied by a factor of $\log(N)$, where $N$ is the total number of quanta in the system.
Intuitively, it can be easily understood that there must be a relation between $\t$ and $\tau$: the thermal distribution is different in the quantum and classical cases, being it either a Bose-Einstein (or a Fermi-Dirac), or a Maxwell-Boltzmann. If a system in the thermodynamic limit is initially out of equilibrium, as time goes by, interactions will drive it towards the corresponding thermal distribution, while in the absence of interactions thermal equilibrium cannot be attained. It is then natural that in the presence of interactions a quantum system and its classical analogue  will differ more and more as they approach equilibrium. In particular, the quantum mechanical treatment allows for the presence of a Bose-Einstein condensate (BEC), while in the classical treatment a BEC cannot form, unless an artificial cutoff is introduced to remove high momentum modes from the theory~\cite{Guth:2014}.

Inspecting Equations~\eqref{1:heisenberg} and~\eqref{1:eom_class}, we can  identify which characteristics make their predictions diverge. The most obvious difference is that Eq.~\eqref{1:heisenberg} is an operator equation while~\eqref{1:eom_class} is not.
The second difference is that Eq.~\eqref{1:heisenberg} allows for the process $l + m \rightarrow j + k$ to happen even if both final states are empty, while such a process is not allowed by Eq.~\eqref{1:eom_class}. 
In the non-interacting case, $\Lambda_{jk}^{lm} = 0$, these differences are not relevant, as both equations are linear and thus the operators $a_j(t)$ and the amplitudes $A_j(t)$ have the same time dependence. This implies that, if $\langle \N_j(t) \rangle = N_j(t)$ initially, it will always be so. In this case $\t$ is infinite.   
Despite its simplicity, the non-interacting case has a wide variety of interesting phenomenology, for example interference, beats, parametric resonance and all the phenomena characteristic to non-interacting  waves. %
On the other hand, if $\Lambda_{jk}^{lm}\neq 0$, both differences are at play. In general, the operator nature of Eq.~\eqref{1:heisenberg} will be mainly responsible for the departure of the quantum description from the classical one (see Ref.~\cite{Sikivie:2016enz}). There is, however, a special set of initial states for which the second difference dominates, at least at initial times. In those states, the distribution of energy is such that the classical evolution cannot proceed, i.e. $\dot{N}_j=0$ for all $j$, on timescales $\lesssim\t$. In this work, we consider such an initial state.

In the present paper, we seek to compare the time evolution of a quantum system governed by the quantum version of the Schr\"odinger-Gross-Pitaevskii (SGP) equation
\begin{equation}
i\partial_t\psi=-\frac{1}{2m}\nabla^2\psi+\frac{\lambda}{8m^2}\psi^\dagger\psi^2  \enspace, \label{eqpsi1}
\end{equation}
with that of a classical system governed by the classical SGP equation 
\begin{equation}
i\partial_t\Psi=-\frac{1}{2m}\nabla^2\Psi+\frac{\lambda}{8m^2}|\Psi|^2\Psi  \enspace, \label{eqPsi1}
\end{equation}
where the field operator $\psi(\vec x,t)$ is replaced by the classical field $\Psi(\vec x,t)$.
Equation~\eqref{eqpsi1} describes, for example, the evolution of a classical real scalar field with contact self-interactions in the non-relativistic limit.
We focus on the case of repulsive contact interactions, $\lambda>0$. 

The quantum treatment of the SGP equation in the case of a homogeneous field at rest was first developed by Bogoliubov~\cite{Bogolyubov:1947zz}. If the interactions are repulsive, such a homogeneous field is stable: the occupation of higher momentum modes does not grow with time.
In this work, we extend Bogoliubov's treatment to the case of an inhomogeneous field.
We consider a particular solution of the linearized classical SGP equation, constituted by a zero-momentum background plus a plane wave perturbation of momentum $\vec{p}$.
As will be shown, this solution gets corrections when the full non-linear classical equation is taken into account, but these corrections only become important on timescales longer than $\t$, unless $p$ is  sufficiently small.
In the quantum description, we find that quanta leave the $\vec{0}$ and  $\vec{p}$ modes in pairs at an exponential rate by parametric resonance, and occupy modes with momentum within an instability window. After a time $\t$, most quanta have jumped out of the $\vec{0}$ and  $\vec{p}$ modes.

Similar calculations were carried out in Ref.~\cite{Chakrabarty:2017fkd} in view of applications to axion dark matter. In Ref.~\cite{Chakrabarty:2017fkd}, the quantum treatment of the initial time evolution of homogeneous fields was given for the cases of attractive contact interactions and for gravitational self-interactions. In both cases, an estimate for $\t$ was provided. This work is the first step in the study of $\t$ for inhomogeneous fields. Here, we find that the introduction of inhomogeneities reduces the duration of classicality from being infinite to being finite. However, a full understanding of the topic requires further work.

Whether quantum correction are important for axion dark matter is still under debate. 
In the literature,  the axion field is usually treated classically~\cite{Sin:1992bg,  Hu:2000ke, Mielke:2009zza, Lora:2011yc, Marsh:2013ywa, Schive:2014dra, Li:2013nal, Hui:2016ltb, Lee:2008jp, Lundgren:2010sp,Marsh:2010wq,RindlerDaller:2011kx,Chavanis:2016dab, Vaquero_2019, Veltmaat:2018dfz}, with few exceptions~\cite{Lentz:2018mpf, Saikawa:2012uk}. Other authors~\cite{Dvali:2017ruz} have investigated the issue of the duration of classicality, using an approach different than ours.

The discussion presented here is also relevant to the description of  Bose-Einstein condensates of ultra-cold atomic gases, whose  interatomic interactions can be modeled as point contact interactions~\cite{burnett1999, pethick_2008}.
It has been experimentally observed (see~\cite{RevModPhys.81.1051} for a review) that, when two such condensates overlap, interference patterns appear.  Based on the results obtained here, we expect that the interference pattern will tend to be smeared out by quantum effects on timescales of order $\t$ because quanta jump to all modes with momentum within the instability window causing the interference pattern to become more blurry. We reserve a detailed discussion of this topic for future work.

This paper is structured as follows. 
In Section~\ref{sec:cl_behavior}, we solve the SGP equation to linear order and estimate under what conditions this solution persists longer than $\t$.
In Section~\ref{sec:q_behavior}, we solve the linearized Heisenberg equations of motion and obtain an analytical expression for $\t$.

\section{Classical behavior} \label{sec:cl_behavior}

Using the notation ${\Psi(\vec x,t)=|\Psi(\vec x,t)|e^{i\theta(\vec x,t)}}$ and defining  the variables
\begin{eqnarray}
n(\vec x, t) &=& |\Psi(\vec x, t)|^2 \label{density}\\
\vec{v}(\vec{x}, t) &=& \frac{1}{m} \vec{\nabla}\theta(\vec{x} ,t)   \enspace, \label{velocity} 
\end{eqnarray}
Eq.~\eqref{eqPsi1} takes the form of a continuity equation
\begin{equation}
 \partial_tn+\vec{\nabla}\cdot(n\vec{v})=0  \label{continuity}
\end{equation}
and an Euler-{\it like} equation
\begin{equation}
 \partial_t\vec{v}+(\vec{v}\cdot\vec{\nabla})\vec{v}=-\frac{1}{m}\vec{\nabla}V-\vec{\nabla}q   \enspace,\label{eulerlike}
\end{equation}
where $V(\vec{x},t)\equiv \frac{\lambda n}{8m^2}$ and $q(\vec{x},t)\equiv \frac{1}{2m^2}\frac{\vec{\nabla}^2\sqrt{n}}{\sqrt{n}}$. 
Hence, the classical field $\Psi(\vec x,t)$  describes a fluid of number density $n$ and velocity $\vec v$. The quantity $q(\vec{x},t)$ is usually called ``quantum pressure'' and distinguishes Eq.~\eqref{eulerlike} from the usual Euler equation for a pressureless perfect fluid.

Equation~(\ref{eqPsi1}) admits the homogeneous solution
\begin{equation}
\Psi_0(t)=\sqrt{n_0}e^{-i\delta\omega t}   \enspace, \label{Psi01}
\end{equation}
where $n_0$ is an $\vec x$-independent number density and
\begin{equation}
\delta\omega=\frac{\lambda n_0}{8m^2}   \enspace. \label{deltaomega1}
\end{equation}
Consider small perturbations about that solution
\begin{equation}
\Psi(\vec x,t)=\Psi_0(t)+\Psi_1(\vec x,t)   \enspace,  \label{Psi0Psi1}
\end{equation}
where $|\Psi_1|\ll|\Psi_0|=\sqrt{n_0}$. Expanding $\Psi_1$ in Fourier modes as
\begin{equation}
\Psi_1(\vec{x}, t) = \Psi_0(t)
\sum_{\vec{k}} C_{\vec{k}}(t) e^{i \vec{k} \cdot \vec{x}}   \enspace,
\label{Four}
\end{equation}
Equation~(\ref{eqPsi1}) gives us
\begin{equation}
i\partial_t C_{\vec k}=\left(\frac{k^2}{2m} - \delta\omega\right)C_{\vec k}
+\delta\omega\sum_{\vec k'\, \vec{k}''} C_{\vec k'}^*\, C_{\vec k''}\, C_{\vec{k} + \vec{k}' - \vec k''}  \enspace. \label{eqC1}
\end{equation}
Equations (\ref{eqC1}) can be solved perturbatively expanding $C_{\vec k}=C_{\vec k}^{(0)}+C_{\vec k}^{(1)}+C_{\vec k}^{(2)} + ...\,$, where $|C_{\vec k}^{(0)}| \gg |C_{\vec k}^{(1)}| \gg |C_{\vec k}^{(2)}|$ and so on. To zeroth order, we have $C^{(0)}_{\vec k}=\delta_{\vec k\, \vec 0}$\,. The first order equation is
\begin{equation}
i\partial_tC_{\vec k}^{(1)}=\left(\frac{k^2}{2m}+\delta\omega\right)C_{\vec k}^{(1)}+\delta\omega C_{-\vec k}^{(1)*}  \enspace. \label{eqC01}
\end{equation}
The solutions may be written as
\begin{equation}
C_{\vec k}^{(1)}(t)=s_{\vec k}(t)+r_{\vec k}(t) \label{Csr}
\end{equation}
with
\begin{equation}
s_{-\vec k}(t)^*=s_{\vec k}(t)\ \ \ \ \text{and}\ \ \ \ r_{-\vec k}(t)^*=-r_{\vec k}(t)  \enspace.  \label{srprop}
\end{equation}
Equations (\ref{eqC01}) imply that
\begin{equation}
r_{\vec k}(t)={2mi\over k^2}\partial_ts_{\vec k}(t)  \label{rsol1}
\end{equation}
and that $s_{\vec k}$ is a solution of
\begin{equation}
\left(\partial_t^2+\omega_k^2\right)s_{\vec k}(t)=0  \enspace, \label{eqs1}
\end{equation}
where
\begin{equation}
\omega_k = \sqrt{{k^2 \over 2m}
\left({k^2 \over 2m} + 2\delta\omega\right)}  \enspace.
\label{omek}
\end{equation}
The most general expression for $s_{\vec k}(t)$ is
\begin{equation}
s_{\vec k}(t)=d_{\vec k}e^{-i\omega_kt}+d_{-\vec k}^*e^{i\omega_kt}  \enspace, \label{sols1}
\end{equation}
where the coefficients $d_{\vec k}$ are determined by the initial density and velocity fields.

\subsection{One mode inhomogeneity}

If the inhomogeneity consists of a one-mode perturbation with momentum $\vec p$, we can write the density field as
\begin{equation}
n(\vec x,t)=n_0+{\delta n\over2}e^{i(\vec p\cdot\vec x-\omega_pt)}+{\delta n^*\over2}e^{-i(\vec p\cdot\vec x-\omega_pt)}   \enspace, \label{pdens}
\end{equation}
where the complex parameter $\delta n$ contains the information about density amplitude and its initial phase. Using equations (\ref{density}), (\ref{Psi0Psi1}) and (\ref{sols1}), one can find that
\begin{equation}
s_{\vec p}(t)={1\over4}{\delta n\over n_0}e^{-i\omega_pt}, \ \ \ \ \ \ \ \ \ \ s_{-\vec p}(t)={1\over4}{\delta n^*\over n_0}e^{i\omega_pt} \label{s2}
\end{equation}
and $s_{\vec k}=0$ for $\vec k\neq\pm\vec p$. We also find
\begin{equation}
r_{\vec p}(t)={1\over4}{\delta n\over n_0}{2m\omega_p\over p^2}e^{-i\omega_pt}, \ \ \ \ \ \ \ \ \ \ r_{-\vec p}(t)=-{1\over4}{\delta n^*\over n_0}{2m\omega_p\over p^2}e^{i\omega_pt} \label{r1}
\end{equation}
and
\begin{equation}
C_{\vec p}^{(1)}(t)={1\over4}{\delta n\over n_0}\left(1+{2m\omega_p\over p^2}\right)e^{-i\omega_pt}, \ \ \ \ \ \ \ \ \ \ C_{-\vec p}^{(1)}(t)={1\over4}{\delta n^*\over n_0}\left(1-{2m\omega_p\over p^2}\right)e^{i\omega_pt}   \enspace. \label{C1}
\end{equation}

Since the solution above is obtained at first order in perturbation theory, the condition $|\Psi_1 | \ll |\Psi_0|$ has to be satisfied.
This is equivalent to
\begin{equation}
 \left|\frac{\delta n}{n_0}\right|\ll\frac{2}{\sqrt{1+4m\delta\omega/p^2}} \sim \frac{p}{\sqrt{m\delta\omega}} \enspace.  
\end{equation}

\subsection{Higher order corrections}
The calculations of Section~\ref{sec:q_behavior} rely on the assumption that the amplitudes $C_{\pm\vec p}^{(1)}$ do not receive significant corrections when the full equations~\eqref{eqC1} are taken into account. To test the validity of this assumption, we investigate the corrections at higher orders in perturbation theory.

We start by noticing that at second order, the amplitudes $C_{\pm 2\vec p}^{(2)}$ and $C_{\vec 0}^{(2)}$ are sourced, at third order the amplitudes $C_{\pm 3\vec p}^{(3)}$ and $C_{\pm \vec p}^{(3)}$ are sourced, and so on. 
Thus, the amplitudes $C_{\pm\vec p}^{(1)}$ receive correction at all odd orders, the main contribution coming from at third order at initial times. 

The third order corrections are given by
\begin{eqnarray}
|C_{\pm\p}^{(3)} |
&\approx&  \left(\frac{|\delta n|}{n_0}\right)^3
\  \delta\omega t 
\ \left|f_\pm \left(\frac{\omega_p}{\delta\omega}\right)\right| \enspace,
\end{eqnarray}
where $f_\pm \left(\frac{\omega_p}{\delta\omega}\right)$ are dimensionless functions.
The amplitudes $|C_{\pm\p}^{(3)} |$ become of the same magnitude as $|C_{\pm\p}^{(1)} |$ over a time
\begin{equation}
t_{\pm\p} \approx 
\frac{1}{\delta\omega }\left(\frac{|\delta n|}{n_0}\right)^{-2}
\frac{\left( 1 \pm \frac{2m\omega_p}{p^2}\right) }{4\left|f_\pm \left(\frac{\omega_p}{\delta\omega}\right)\right|}   \enspace.
\end{equation}

\begin{figure}[ht]
     \centering
     \includegraphics[width=\textwidth]{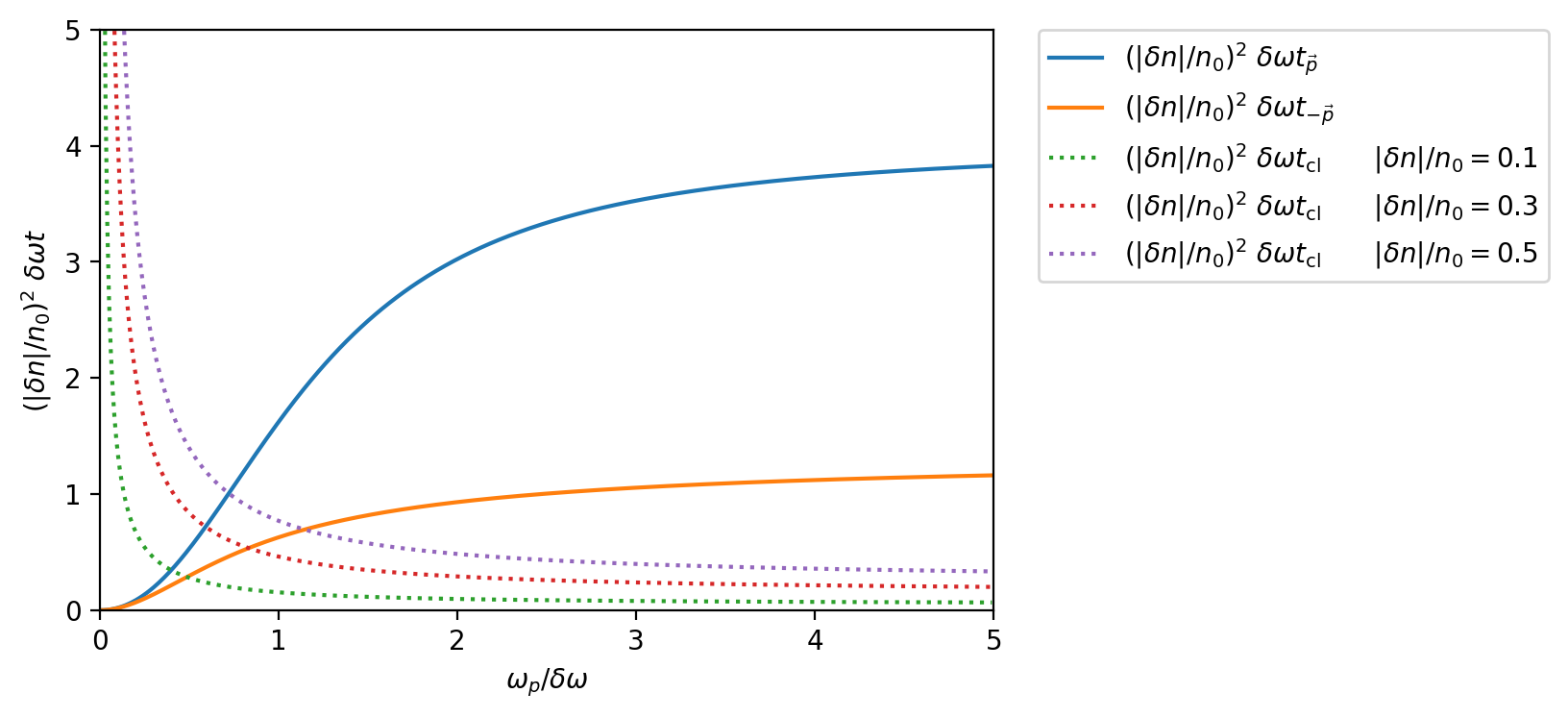}
     \caption{Comparison of $t_{\pm\p}$ and $t_{\rm cl}$ as a function of $\omega_p/\delta\omega$ for various value of the density contrast $|\delta n|/n_0$. $t_{\rm cl}$ is obtained from Eq.~\eqref{t_classicality} setting the logarithm factor to 1. For given $|\delta n|/n_0$, the calculations of Section \ref{sec:q_behavior} are valid for all values of $\omega_p$ such that the corresponding dotted line lies lower than the solid ones.}
     \label{scales}
\end{figure}

Neglecting the logarithmic factor in Eq.~\eqref{t_classicality}, the classical third order corrections grow more slowly then the quantum corrections, except if 
\begin{equation}
\omega_p \lesssim  \delta\omega \left(  \frac{2}{3}\frac{|\delta n|}{n_0} \right)^{1/3} \enspace.
\end{equation} 
In Figure~\ref{scales}, we show how $t_{\pm\p}\,$ compare to  the duration of classicality Eq.~\eqref{t_classicality}. 
We checked numerically that higher order corrections do not play a significant role of timescales of order $\t$.

\section{Quantum Corrections}\label{sec:q_behavior}

In order to study the quantum corrections to the initial classical description, we write
\begin{equation}
\psi(\vec x,t)=\Psi(\vec x,t)+\phi(\vec x,t)  \enspace, \label{phidef}
\end{equation}
where $\Psi$ is a solution of the classical SGP equation and $\phi$ an operator containing all the information about quantum corrections. 
Initially, $|\Psi| \sim \sqrt{N}$, where $N$ is the total number of quanta in the system, while $\phi \approx \mathcal{O}(1)$. 
Replacing (\ref{phidef}) into (\ref{eqpsi1}) and keeping the leading term in an expansion in powers of $1/\sqrt{N}$, we find
\begin{equation}
i\partial_t\phi=-{1\over2m}\nabla^2\phi+{\delta\omega\over n_0}\left(2|\Psi|^2\phi+\Psi^2\phi^\dagger\right)  \enspace. \label{eqphi1}
\end{equation}
We expand $\phi$ in the form
\begin{equation}
\phi(\vec x,t)={e^{-i\delta\omega t}\over\sqrt{V}}\sum_{\vec k}b_{\vec k}(t)e^{i\vec k\cdot\vec x}  \enspace, \label{phiexp}
\end{equation}
where $b_{\vec k}$ are time dependent operators  satisfying the canonical commutation relations
\begin{equation}
[b_{\vec k}(t),b_{\vec k'}(t)]=0, \ \ \ \ \ \ \ \ \ \ [b_{\vec k}(t),b_{\vec k'}(t)^\dagger]=\delta_{\vec k}^{\vec k'}
\end{equation}
and $V$ is the volume of space where the theory is defined.
The equations of motion for $b_{\vec k}(t)$ are
\begin{equation}
i\partial_tb_{\vec k}=a_kb_{\vec k}+\delta\omega b_{-\vec k}^\dagger+2\delta\omega\sum_{\vec k'}\left(2s_{\vec k'}b_{\vec k-\vec k'}+C_{\vec k'}b_{\vec k'-\vec k}^\dagger\right)  \enspace, \label{eqb1}
\end{equation}
where $a_k=k^2/2m+\delta\omega$.
We perform a Bogoliubov transformation
\begin{equation}
\left(
\begin{array}{cc}
b_{\vec k}
\\
b_{-\vec k}^\dagger
\end{array}
\right) = \left(
\begin{array}{cc}
u_k & v_k
\\
v_k & u_k
\end{array}
\right) \left(
\begin{array}{cc}
\beta_{\vec k}
\\
\beta_{-\vec k}^\dagger
\end{array}
\right)  \enspace, \label{btobeta}
\end{equation}
where $u_k$ and $v_k$ are real and $u_k^2-v_k^2=1$ is required for the transformation from the $b_{\vec k}$ to the $\beta_{\vec k}$ to be canonical. We  may write $u_k=\cosh(\eta_k)$ and $v_k=\sinh(\eta_k)$. Choosing $\eta_k$ such that $\tanh(2\eta_k)=-\delta\omega/a_k$, Eq. (\ref{eqb1}) becomes
\begin{equation}
i\partial_t\beta_{\vec k}=\omega(k)\beta_{\vec k}+2\delta\omega\sum_{\vec k'}\left[{\cal P}_{\vec k}^{\vec k'}(t)\beta_{\vec k-\vec k'}+{\cal Q}_{\vec k}^{\vec k'}(t)\beta_{-\vec k+\vec k'}^\dagger\right]   \enspace, \label{eqbeta1}
\end{equation}
where
\begin{equation}
{\cal P}_{\vec k}^{\vec k'}(t)=2s_{\vec k'}(t)(u_ku_{|\vec k-\vec k'|}+v_kv_{|\vec k-\vec k'|})+C_{\vec k'}(t)u_kv_{|\vec k-\vec k'|}+C_{-\vec k'}(t)^*v_ku_{|\vec k-\vec k'|} \label{Pcal1}
\end{equation}
and
\begin{equation}
{\cal Q}_{\vec k}^{\vec k'}(t)=2s_{\vec k'}(t)(u_kv_{|\vec k-\vec k'|}+v_ku_{|\vec k-\vec k'|})+C_{\vec k'}(t)u_ku_{|\vec k-\vec k'|}+C_{-\vec k'}(t)^*v_kv_{|\vec k-\vec k'|} \enspace. \label{Qcal1}
\end{equation}

\subsection{One mode inhomogeneity}
\begin{figure}[h!]
\begin{center}
    \includegraphics[width=\linewidth]{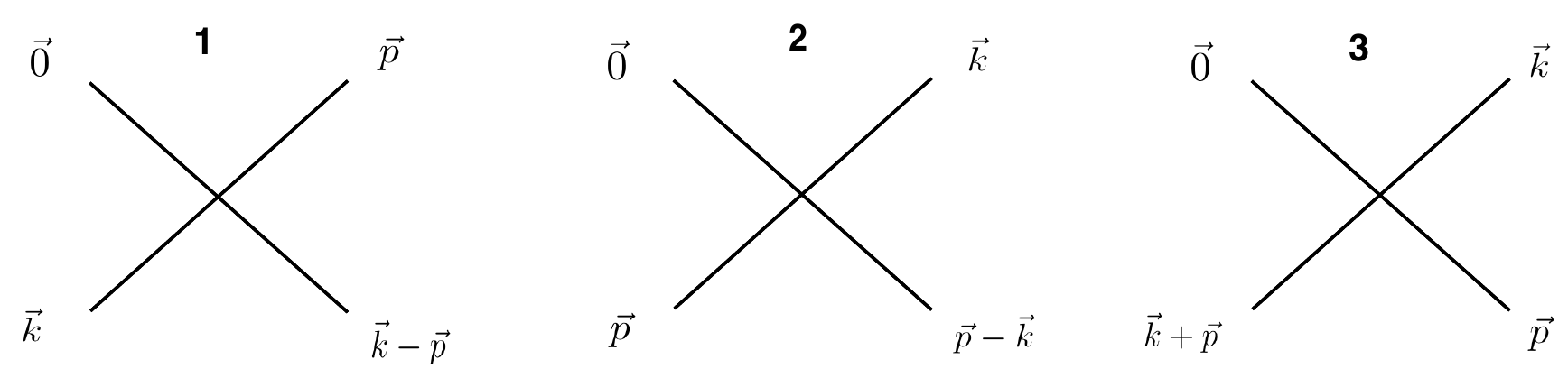}
\caption{Feynmann diagrams of the processes.}
\label{diagrams}
\end{center}
\end{figure}

We specialize to the simple case in which the perturbation has a definite momentum $\vec{p}$. The only non-zero terms in the sum of Eq. (\ref{eqbeta1}) are those proportional to $s_{\vec{p}}(t)$, $s_{-\vec{p}}(t)$, $C_{\vec{p}}(t)$ and $C_{-\vec{p}}(t)$. Eq. (\ref{eqbeta1})  becomes
\begin{eqnarray}
i\partial_t\beta_{\vec k}-\omega(k)\beta_{\vec k}  &=& \frac{\delta\omega}{2}\frac{\delta n}{n_0}e^{-i\omega_pt}\left(P_{\vec k}^{\vec p\ (+)}\beta_{\vec k-\vec p}+Q_{\vec k}^{\vec p\ (+)}\beta_{-\vec k+\vec p}^\dagger\right) \nonumber
\\
& &+\frac{\delta\omega}{2}\frac{\delta n^*}{n_0}e^{i\omega_pt}\left(P_{\vec k}^{\vec p\ (-)}\beta_{\vec k+\vec p}+Q_{\vec k}^{\vec p\ (-)}\beta_{-\vec k-\vec p}^\dagger\right)  \enspace, \label{eqbeta2}
\end{eqnarray}
where
\begin{equation}
P_{\vec k}^{\vec p\ (\pm)}=2u_ku_{|\vec k\mp\vec p|}+2v_kv_{|\vec k\mp\vec p|}+u_kv_{|\vec k\mp\vec p|}+v_ku_{|\vec k\mp\vec p|}\pm{2m\omega_p\over p^2}(u_kv_{|\vec k\mp\vec p|}-v_ku_{|\vec k\mp\vec p|}) \label{P1}
\end{equation}
and
\begin{equation}
Q_{\vec k}^{\vec p\ (\pm)}=2u_kv_{|\vec k\mp\vec p|}+2v_ku_{|\vec k\mp\vec p|}+u_ku_{|\vec k\mp\vec p|}+v_kv_{|\vec k\mp\vec p|}\pm{2m\omega_p\over p^2}(u_ku_{|\vec k\mp\vec p|}-v_kv_{|\vec k\mp\vec p|})  \enspace. \label{Q1}
\end{equation}
They satisfy the identities
\begin{equation}
P_{\vec k-\vec p}^{\vec p\ (-)}=P_{\vec k}^{\vec p\ (+)}, \ \ \ \ \ \ \ \ \ \ Q_{-\vec k+\vec p}^{\vec p\ (+)}=Q_{\vec k}^{\vec p\ (+)}  \enspace, \label{PQid}
\end{equation}
which will be used later. Writing $\beta_{\vec k}=\alpha_{\vec k}\ e^{-i\omega_kt}$, we have
\begin{eqnarray}
i\partial_t \alpha_{\vec k}  &=& \frac{\delta\omega}{2}\frac{\delta n}{n_0}\left(P_{\vec k}^{\vec p\ (+)}\alpha_{\vec k-\vec p}\ e^{-i\delta_{\vec k}^{\vec p}t}+Q_{\vec k}^{\vec p\ (+)}\alpha_{-\vec k+\vec p}^\dagger\ e^{-i\epsilon_{\vec k}^{\vec p}t}\right)
\\
& &+\frac{\delta\omega}{2}\frac{\delta n^*}{n_0} \left(P_{\vec k}^{\vec p\ (-)}\alpha_{\vec k+\vec p}\ e^{-i\phi_{\vec k}^{\vec p}t} +Q_{\vec k}^{\vec p\ (-)}\alpha_{-\vec k - \vec p}^\dagger \ e^{-i\gamma_{\vec k}^{\vec p}t}\right) \enspace, 
\end{eqnarray}

where
\begin{eqnarray}
\delta_{\vec k}^{\vec p} &=& - \omega_k + \omega_p + \omega_{|\vec k - \vec p|}  \enspace, \nonumber 
\\
\epsilon_{\vec k}^{\vec p} &=& - \omega_k +\omega_p - \omega_{|\vec k - \vec p|}  \enspace, \nonumber 
\\
\phi_{\vec k}^{\vec p} &=& - \omega_k  -\omega_p  + \omega_{|\vec k + \vec p|}  \enspace, \nonumber
\\
\gamma_{\vec k}^{\vec p} &=& - \omega_k -\omega_p  - \omega_{|\vec k + \vec p|} \enspace.
\end{eqnarray}
When $\delta_{\vec k}^{\vec p}$, $\epsilon_{\vec k}^{\vec p}$ or $\phi_{\vec k}^{\vec p}$ approach 0, some scattering processes get excited which are shown in Fig.~(\ref{diagrams}). $\gamma_{\vec k}^{\vec p}=0$ is prohibited by conservation of energy. In our initial state, only the quasi-particles states of momentum $\vec{0}$ and $\vec{p}$ are occupied. Therefore, as we show below, only process 2 is truly important.

\subsection{Parametric instability}

For process 1, the relevant equations are
\begin{eqnarray}
i\partial_t\alpha_{\vec{k}} &\approx & \frac{\delta\omega}{2}\frac{\delta n}{n_0}P_{\vec{k}}^{\vec{p}\ (+)} \alpha_{\vec{k}-\vec{p}}\ e^{-i\delta_{\vec k}^{\vec p}t}	\label{eqe11}
\\
i\partial_t\alpha_{\vec{k}-\vec{p}} &\approx &\frac{\delta\omega}{2}\frac{\delta n^*}{n_0}P_{\vec{k}}^{\vec{p}\ (+)} \alpha_{\vec{k}}\ e^{i\delta_{\vec k}^{\vec p}t} \label{eqe12}   \enspace,
\end{eqnarray}
which we have found using (\ref{PQid}) and the property $\phi_{\vec k-\vec p}^{\vec p}=-\delta_{\vec k}^{\vec p}$. Defining $X_{\vec k}=\alpha_{\vec k}\,e^{{i\over2}\delta_{\vec k}^{\vec p}t}$ and $Y_{\vec k}=\alpha_{\vec k-\vec p}\,e^{-{i\over2}\delta_{\vec k}^{\vec p}t}$, Equations (\ref{eqe11}) and (\ref{eqe12}) become
\begin{equation}
\left[\partial_t^2+\left({\delta\omega\over 2}{|\delta n|\over n_0}P_{\vec k}^{\vec p\ (+)}\right)^2+{1\over4}{\delta_{\vec k}^{\vec p}}^{\ 2}\right]\left[
\begin{array}{cc}
X_{\vec k}
\\
Y_{\vec k}
\end{array}
\right]=0  \enspace. \label{e1osc}
\end{equation}
In this case, $\alpha_{\vec k}$ and $\alpha_{\vec k - \vec{p}}$  oscillate implying that the number of quanta occupying these modes does not grow with time. Such a process is not significant for our purposes. For process 3, the equations have the same form as those in process 1 and we have trivial oscillations in this case as well.

\begin{figure}[!ht]
\begin{center}
    \includegraphics[width=\linewidth]{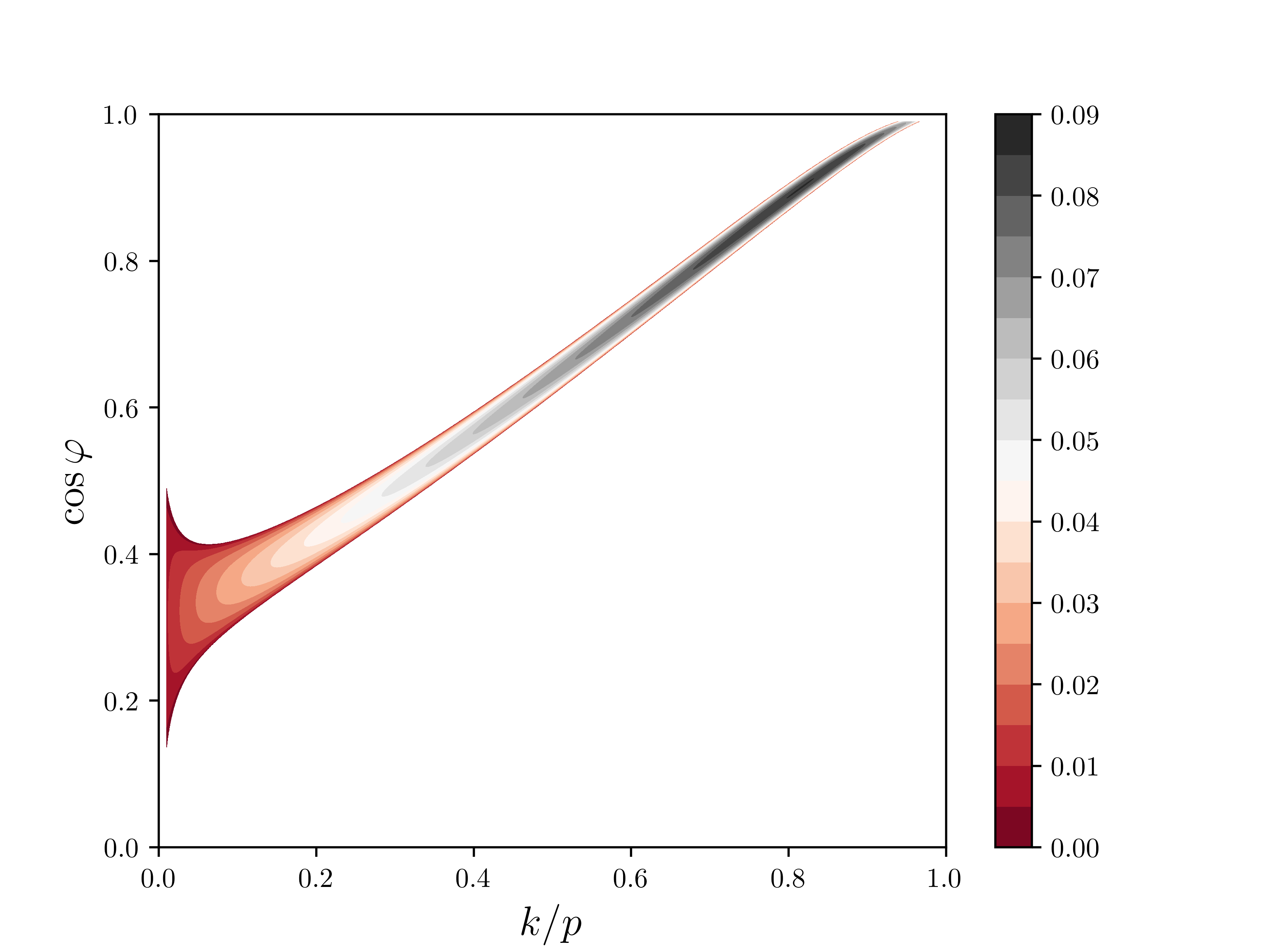}
\caption{Contour plot of $S(\vec{k},\vec{p}) /\delta\omega$, for $|\delta n|/ n_0 = 0.1$ and $\omega_p/\delta\omega = 6$. The horizontal axis is the ratio of the momentum magnitudes $k/p$, while the vertical axis is the cosine of the angle between $\vec{k}$ and $\vec{p}$.}
\label{contour}
\end{center}
\end{figure}

For process 2, the relevant equations are
\begin{eqnarray}
i\partial_t\alpha_{\vec{k}} &\approx & \frac{\delta\omega}{2}\frac{\delta n}{n_0} Q_{\vec k}^{\vec p\ (+)}\alpha^{\dagger}_{\vec{p}-\vec{k}}\ e^{-i\epsilon_{\vec k}^{\vec p}t} \label{eqe21}
 \\
-i\partial_t\alpha^{\dagger}_{\vec{p}-\vec{k}} &\approx &\frac{\delta\omega}{2}\frac{\delta n^*}{n_0}Q_{\vec k}^{\vec p\ (+)}\alpha_{\vec{k}}\ e^{i\epsilon_{\vec k}^{\vec p}t} \enspace, \label{eqe22}
\end{eqnarray}
which we have found using (\ref{PQid}) and the property $\epsilon_{-\vec k+\vec p}^{\vec p}=\epsilon_{\vec k}^{\vec p}$. Defining $\alpha_{\vec k}(t)=\rho_{\vec k}(t)e^{-{i\over2}\epsilon_{\vec k}^{\vec p}t}$, Equations (\ref{eqe21}) and (\ref{eqe22}) become
\begin{equation}
\left[\partial_t^2-\left({\delta\omega\over 2}{|\delta n|\over n_0}Q_{\vec k}^{\vec p\ (+)}\right)^2+{1\over4}{\epsilon_{\vec k}^{\vec p}}^{\ 2}\right]\left[
\begin{array}{cc}
\rho_{\vec k}
\\
\rho_{\vec p-\vec k}^\dagger
\end{array}
\right]=0  \enspace. \label{e2res}
\end{equation}
The equations above describe parametric resonance in the neighbourhood of $\epsilon_{\vec k}^{\vec p}= 0$. The condition for the resonance is
\begin{equation}
-\delta\omega{|\delta n|\over n_0}Q_{\vec k}^{\vec p}<\epsilon_{\vec k}^{\vec p}<\delta\omega{|\delta n|\over n_0}Q_{\vec k}^{\vec p}  \enspace, \label{rescon}
\end{equation}
where we have defined $Q_{\vec k}^{\vec p}=Q_{\vec k}^{\vec p\ (+)}$. 
If $\vec k$ is such that the condition~(\ref{rescon}) is satisfied, $\rho_{\vec{k}}(t)$ grows exponentially at a rate
\begin{equation}
S_{\vec k}^{\vec p}={1\over 2}\sqrt{\left(\delta\omega{|\delta n|\over n_0}Q_{\vec k}^{\vec p}\right)^2-{\epsilon_{\vec k}^{\vec p}}^{\ 2}} \label{S1} \enspace.
\end{equation}
The region of instability is shown in Figure~\ref{contour} for sample values of the parameters.

\subsection{Duration of classicality}
Equations (\ref{eqe21}) and (\ref{eqe22}) can be written as
\begin{equation}
i\partial_t\left(
\begin{array}{cc}
\rho_{\vec k}
\\
\rho_{\vec p-\vec k}^\dagger
\end{array}
\right)
={1\over 2}\left(
\begin{array}{cc}
-\epsilon_{\vec k}^{\vec p} & q_{\vec k}^{\vec p}
\\
-{q_{\vec k}^{\vec p}}^{\ *} & \epsilon_{\vec k}^{\vec p}
\end{array}
\right)\left(
\begin{array}{cc}
\rho_{\vec k}
\\
\rho_{\vec p-\vec k}^\dagger
\end{array}
\right)  \enspace, \label{rhok-p}
\end{equation}
where $q_{\vec k}^{\vec p}={\delta n\over n_0}\delta\omega Q_{\vec k}^{\vec p}$. 
We perform another Bogoliubov transformation 
\begin{equation}
\left(
\begin{array}{cc}
\rho_{\vec k}
\\
\rho_{\vec p-\vec k}^\dagger
\end{array}
\right) = \left(
\begin{array}{cc}
w & z
\\
z^* & w^*
\end{array}
\right) \left(
\begin{array}{cc}
\tilde\rho_{\vec k}
\\
\tilde\rho_{\vec p-\vec k}^\dagger
\end{array}
\right)   \enspace. \label{rhotorhotilde}
\end{equation}
Writing $w\equiv\cosh(\tilde\eta_{\vec k}^{\vec p})e^{i\theta/2}$, $z\equiv\sinh(\tilde\eta_{\vec k}^{\vec p})e^{i\theta/2}$, $q_{\vec k}^{\vec p}=|q_{\vec k}^{\vec p}|e^{i\theta}$ and choosing $\tilde\eta_{\vec k}^{\vec p}$ such that $\tanh(2\tilde\eta_{\vec k}^{\vec p})=\epsilon_{\vec k}^{\vec p}/|q_{\vec k}^{\vec p}|$, Equation (\ref{rhok-p}) becomes
\begin{equation}
i\partial_t\left(
\begin{array}{cc}
\tilde\rho_{\vec k}
\\
\tilde\rho_{\vec p-\vec k}^\dagger
\end{array}
\right)
=\left(
\begin{array}{cc}
0 & S_{\vec k}^{\vec p}
\\
-S_{\vec k}^{\vec p} & 0
\end{array}
\right)\left(
\begin{array}{cc}
\tilde\rho_{\vec k}
\\
\tilde\rho_{\vec p-\vec k}^\dagger
\end{array}
\right)  \enspace. \label{eqrhotilde1}
\end{equation}
The solution of (\ref{eqrhotilde1}) is 
\begin{equation}
\tilde\rho_{\vec k}(t)=\tilde\rho_{\vec k}(0)\cosh(S_{\vec k}^{\vec p}t)-i\tilde\rho_{\vec p-\vec k}^\dagger(0)\sinh(S_{\vec k}^{\vec p}t)  \enspace. \label{solrhotilde1}
\end{equation}

To compute expectation values, we choose the state of the system $\left|\Psi\right>$ as the following
\begin{equation}
\tilde{\rho}_{\vec{k}}(0)\left|\Psi\right> = \tilde{\rho}_{\vec{p} - \vec{k}}(0)\left|\Psi\right> = 0  \enspace. \label{initial}
\end{equation}
Based on the findings of Ref.~\cite{Chakrabarty:2017fkd}, we expect the state in Eq.~(\ref{initial}) to have the longest duration of classicality. The occupation number of a mode with $\vec{k} \neq 0$ is given by
\begin{eqnarray}
\left< N_{\vec{k}}(t) \right> &=& \left<\Psi \right|b_{\vec k}^\dagger(t)b_{\vec k}(t)\left|\Psi\right> \nonumber \\ 
&=& \cosh(\eta_{\vec k})^2\left( \sinh(\tilde\eta_{\vec k}^{\vec p})^2 \cosh(S_{\vec k}^{\vec p}t)^2 + \cosh(\tilde\eta_{\vec k}^{\vec p})^2\sinh(S_{\vec k}^{\vec p}t)^2\right) + \frac{1}{2} \sinh(\eta_{\vec k})^2 . \label{occupation_num}
\end{eqnarray}
Notice that $S_{\vec p}^{\vec p} = 0$.
After sufficiently long time such that $S_{\vec k}^{\vec p}t \gg 1$, 
\begin{equation}
\left< N_{\vec{k}}(t) \right> \approx \frac{1}{16} \left( \frac{a_k}{\omega_k} + 1 \right) \frac{|q_{\vec k}^{\vec p}|}{S_{\vec k}^{\vec p}} e^{2S_{\vec k}^{\vec p}t}   \enspace.\label{occupation_num_long}
\end{equation}
After time $t$, the total number of quanta that have left the $\vec{0}$ and $\vec{p}$ modes corresponding to the classical solution is 
\begin{eqnarray}
N_{\text{ev}}(t) = \sum_{\vec{k} \neq 0} \left< N_{\vec{k}}(t) \right> = \frac{V}{4\pi^2} \int k^2 dk\, d\mu \left< N_{\vec{k}}(t) \right>   \enspace,\label{N_ev}
\end{eqnarray}
where $\mu = \cos \phi$ with $\phi$ being the angle between $\vec{k}$ and $\vec{p}$. To perform the integral, we change the variables from $(k, \mu)$ to $(\omega_1, \omega_2)$ where $\omega_1 = \omega_k$ and $\omega_2 = \omega_{|\vec k-\vec p|}$. Then we have 
\begin{equation}
N_{\text{ev}}(t) = \frac{V}{4\pi^2} \int d\omega_1 d\omega_2 \; 2m \left( \sqrt{\omega_1^2 + \delta \omega^2} - \delta \omega \right) \left| \frac{\partial(k, \mu)}{\partial(\omega_1, \omega_2)} \right| \left< N_{\vec{k}}(t) \right>   \enspace.\label{N_ev_omega}
\end{equation}
Since $\left< N_{\vec{k}} (t) \right>$ grows exponentially (see Eq.~(\ref{occupation_num_long})), we use saddle-point approximation. $S_{\vec k}^{\vec p}$ has a maximum when $\omega_1 = \omega_2$ and $\epsilon_{\vec k}^{\vec p} \approx 0$. This leads to $\omega_1 = \omega_2 = \frac{\omega_p}{2}$. After tedious but straightforward calculations, we obtain 
\begin{equation}
    N_{\text{ev}}(t) = \frac{N}{n_0} \ \frac{1}{\delta \omega t} \ (m\delta \omega)^{\frac{3}{2}}  \ F(\Lambda) \ e^{Q(\Lambda) \delta\omega t \left(\frac{|\delta n|}{n_0}\right)} \enspace, \label{N_ev_final}
\end{equation}
where $\Lambda=\omega_p/\delta\omega$ and $F(\Lambda)$ is a dimensionless function of $\Lambda$ given in the Appendix and shown in Fig.~\ref{FLambda}. $Q(\Lambda)$ is given by
\begin{equation}
Q(\Lambda) = \frac{1}{\Lambda} \left( \sqrt{\Lambda^2 + 4} + \sqrt{\Lambda^2 + 1} - 3 \right) \ . \label{QLambda_repulsive}
\end{equation}
The duration of classicality $t_{\text{cl}}$ is the time $t$ such that $N_{\text{ev}}(t) \approx N$:
\begin{eqnarray}
t_{\text{cl}} &\approx& \frac{1}{Q(\Lambda) \ \delta \omega \ \left(\frac{|\delta n|}{n_0}\right)} \ \ln \left( \frac{1}{F(\Lambda)} \ \frac{n_0}{(m \delta \omega)^{\frac{3}{2}}} \right) \enspace. \label{t_classicality}
\end{eqnarray}
The argument of the logarithm is proportional to the number of particles in a volume $(m \delta \omega)^{-3/2}$.

In the high and low momentum limits, we obtain
\begin{eqnarray}
t_{\text{cl}} &\approx&  \left\{\begin{array}{ll} \displaystyle \frac{1}{2}\cdot\frac{1}{\delta \omega \ \frac{|\delta n|}{n_0}} \ \ln \left( 128\sqrt{2}\pi\frac{m\delta\omega}{p^2} \cdot \frac{n_0}{(m \delta \omega)^{\frac{3}{2}}} \right) 
& \quad\mathrm{for}\quad \frac{p}{\sqrt{m\delta\omega}}\gg 1 \\ 
\displaystyle \frac{4\sqrt{m\delta\omega}}{3p}\cdot \frac{1}{\delta \omega \ \frac{|\delta n|}{n_0}} \ln \left(128\pi \frac{\sqrt{m\delta\omega}}{p} \cdot \frac{n_0}{(m \delta \omega)^{\frac{3}{2}}} \right) 
& \quad\mathrm{for}\quad \left|\frac{\delta n}{n_0}\right|\ll\frac{p}{\sqrt{m\delta\omega}}\ll 1 \enspace.
\end{array}\right. 
\end{eqnarray}
This result shows us that duration of classicallity is inversely proportional to the density contrast, and that the smaller the momentum $p$, the longer the duration of classicality.

\begin{figure}[h!]
\begin{center}
    \includegraphics[scale=0.7]{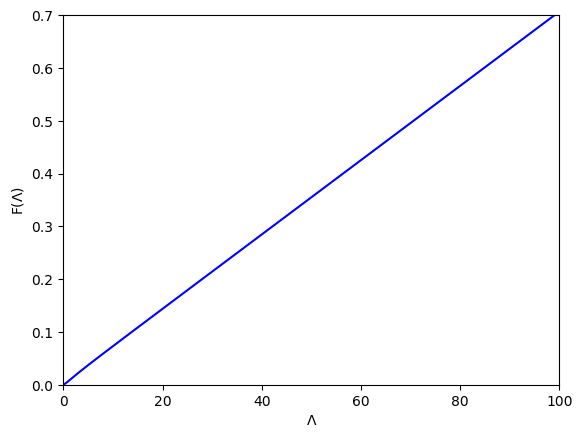}
\caption[Plot of $F(\Lambda)$]{Plot of $F(\Lambda)$ which is a dimensionless function of $\Lambda = \omega_p/\delta\omega $.}
\label{FLambda}
\end{center}
\end{figure}

\section{Summary and Conclusions}\label{sec:concl}
This work was motivated by the following question: ``how long can a highly degenerate quantum scalar field be described accurately by classical field equations?''. Intuitively, this time scale cannot be longer than the thermalization time scale. A generic formalism to calculate the duration of classicality was developed in Ref.~\cite{Chakrabarty:2017fkd}. There, the formalism was applied to a homogeneous field with attractive contact interactions or gravitational self-interactions. Classically, the homogeneous state persists forever. However, in the quantum evolution, small quantum fluctuations grow exponentially due to attractive nature of the interactions and the homogeneous state gets depleted~\cite{Chakrabarty:2017fkd}. 

In this work, we have considered the case of repulsive contact self-interactions. We have focused on an inhomogeneous solution of the classical equations of motion made by a zero momentum background and a small plane wave perturbation of momentum $\vec{p}$.As shown in Sec.~\ref{sec:cl_behavior}, in the classical description, this solution is stable up to third order corrections that play an important role only at low $p$. In Sec.~\ref{sec:q_behavior}, we have studied the system using quantum field theory. The quantum evolution becomes illuminating in terms of operator $\beta_{\vec{k}}$ which annihilates a quasi-particle of momentum $\vec{k}$ and energy $\omega(\vec{k})$ (see Eq.~(\ref{omek}) and~(\ref{eqbeta1})). Classically, the quasi-particles remain in the $\vec{0}$ and $\vec{p}$ modes. In the quantum description, the quasi-particles scatter through the process $\vec{0} + \vec{p} \rightarrow \vec{k} + (\vec{p} - \vec{k})$. This process is enhanced by parametric resonance if $\vec{k}$ lies on the surface in momentum space defined by $\omega(\vec{0}) + \omega(\vec{p}) = \omega(\vec{k}) + \omega(\vec{p} - \vec{k})$. We have determined a region of instability around this surface. The modes with $\vec{k}$ outside this region of instability are never populated by parametric resonance. Finally, we have estimated the duration of classicality (see Eq.~(\ref{t_classicality})), after which almost all the quasi-particles have left the $\vec{0}$ and $\vec{p}$ modes and the classical description is invalid.

\begin{acknowledgments}

We thank Pierre Sikivie for many useful discussions. This work was supported in part by the U.S. Department of Energy under grant DE-SC0010296 and by the Heising-Simons Foundation under grant No. 2015-109. A.A. was supported in part by the Chilean Commission on
Research, Science and Technology (CONICYT) under grant 78180100 (Becas Chile, Postdoctorado). S.C. was supported in part by the Dissertation Fellowship by the College of Liberal Arts and Sciences, University of Florida. S.E. was supported by the Sun Yat-Sen
University Science Foundation and by JSPS KAKENHI Grants No. JP18H03708 and No. JP17H01131. E.T. is supported by the European Union’s Horizon 2020 research and innovation programme under the Marie Sklodowska-Curie grant agreement No 674896 (Elusives).

\end{acknowledgments}

\appendix
\section{}
Here we provide the dimensionless function $F(\Lambda)$ in Eq.~\ref{N_ev_final} with $\Lambda = \frac{\omega_p}{\delta \omega}$:
\begin{equation}
F(\Lambda) = \frac{1}{4\sqrt{2} \pi} \ f(\Lambda) \ \sqrt{\frac{Q(\Lambda)}{G(\Lambda)}}
\end{equation}
where
\begin{eqnarray}
f(\Lambda) &=& \frac{\Lambda \left( \Lambda + \sqrt{\Lambda^2 + 4} \right)}{8 \left( \Lambda^2 + 4 \right) \ \sqrt{\sqrt{\Lambda^2 + 1} -1}}
\sim \left\{ \begin{array}{lc}\displaystyle\frac{1}{8\sqrt{2}}\left(1+\frac{1}{2}\Lambda+\cdots\right) & \quad(\Lambda\ll 1)\\ \displaystyle\frac{1}{4\sqrt{\Lambda}}\left(1+\frac{1}{2\Lambda}+\cdots\right) & \quad(\Lambda\gg 1) \end{array}\right. ,\nonumber \\
Q(\Lambda) &=& \frac{1}{\Lambda} \left( \sqrt{\Lambda^2 + 4} + \sqrt{\Lambda^2 + 1} - 3 \right) \sim \left\{ \begin{array}{cc}\displaystyle\frac{3}{4}\Lambda\left(1-\frac{3}{16}\Lambda^2+\cdots\right) & (\Lambda\ll 1) \\ \displaystyle 2\left(1-\frac{3}{2\Lambda}+\cdots\right) & (\Lambda\gg 1) \end{array}\right. , \nonumber \\
G(\Lambda) &=& \frac{16}{\Lambda^3 \ \left(\Lambda^2 + 4\right)^{\frac{3}{2}}} \ \left( \sqrt{\Lambda^2 + 4} \left(2 \Lambda^2 + 3 - \sqrt{\Lambda^2 + 1}\right) - 2\left(\Lambda^2 + 2\right) \right)\nonumber\\
 &\sim& \left\{ \begin{array}{cc}\displaystyle\frac{3}{\Lambda}\left(1+\frac{1}{48}\Lambda^2+\cdots\right) & \quad (\Lambda\ll 1), \\ \displaystyle\frac{32}{\Lambda^3}\left(1-\frac{3}{2\Lambda}+\cdots\right) & \quad (\Lambda\gg 1)\end{array}\right. \ \ .
\end{eqnarray}
In the limit of $\Lambda\ll 1$ and $\Lambda\gg 1$, the function $F(\Lambda)$ behaves as
\begin{equation}
 F(\Lambda) \sim \frac{\Lambda}{128\pi} \times \left\{ \begin{array}{cc} 1 & \quad (\Lambda\ll 1), \\ \sqrt{2} & \quad (\Lambda\gg 1)\end{array}\right. \ \ .
\end{equation}
In both cases, $F(\Lambda)$ behaves as a linear function.
The function $F(\Lambda)$ is plotted in Fig.~\ref{FLambda}.



\bibliographystyle{apsrev4-1}
\bibliography{refs.bib}



\end{document}